\documentclass[usenatbib,twocolumn]{mn2e}
\usepackage{times}
\usepackage{graphicx}
\usepackage{amsfonts}
\usepackage{epsfig,rotating,natbib,pstricks}
\usepackage{color}
\usepackage{amssymb,latexsym}
\usepackage{amsmath}

\pubyear{2010}

\def\beq{\begin{equation}}
\def\eeq{\end{equation}} 
\def\baq{\begin{eqnarray}}
\def\eaq{\end{eqnarray}}

\def\p3m{P$^3$M}
\def\ap3m{AP$^3$M}

\def\h1{H\/I}
\def\omegah1{\Omega_{\h1}}
\def\ph1{P_{_{\h1}}}
\def\ph1k{P_{_{\h1}}(k)}
\def\dh1k{\Delta^2_{_{\h1}}(k)}
\def\mmin{M_{min}}
\def\mmax{M_{max}}
\def\msun{M_{\odot}}

\newcommand{\be}{\begin{equation}}
\newcommand{\e}{\end{equation}}
\newcommand{\f}{\frac}
\newcommand{\U}{{{\mathbf U}}}
\newcommand{\Th}{{{\mathbf{\Theta}}}}

\def\mnras{MNRAS}

\title[H\/I in the Post-Reionization Universe]{H\/I as a Probe of the Large Scale Structure in the
  Post-Reionization Universe}

\author[Bagla, Khandai \& Datta]
{J. S. Bagla$^1$, Nishikanta Khandai$^{1,2}$, Kanan K. Datta$^{1,3,4}$   \\
$^1$ Harish-Chandra Research Institute, Chhatnag Road, Jhunsi, 
Allahabad 211019, INDIA \\
$^2$ Department of Physics, Carnegie Mellon University, Pittsburgh, PA 15213, U.S.A.\\
$^3$ Department of Physics and Meteorology \& Centre for Theoretical Studies,
IIT, Kharagpur 721302, India\\
$^4$ The Oskar Klein Centre for Cosmoparticle Physics,
Department of Astronomy, Stockholm University,
Albanova, SE-10691 Stockholm, Sweden\\
E-mail: jasjeet@hri.res.in, nkhandai@andrew.cmu.edu, kdatt@astro.su.se}
\pubyear{2010}

\label{firstpage}

\def\LaTeX{L\kern-.36em\raise.3ex\hbox{a}\kern-.15em
    T\kern-.1667em\lower.7ex\hbox{E}\kern-.125emX}

\pagerange{\pageref{firstpage}--\pageref{lastpage}} 

\begin{document}

\maketitle

\begin{abstract}
We model the distribution of neutral Hydrogen (H\/I hereafter) in the
post-reionization universe. 
This model uses gravity only N-Body simulations and an ansatz to assign
H\/I to dark matter haloes that is consistent with observational
constraints and theoretical models. 
We resolve the smallest haloes that are likely to host \h1 in the
simulations, care is also taken to ensure that any errors due to the finite
size of the simulation box are small.
We then compute the smoothed one point probability distribution function and
the power spectrum of fluctuations in H\/I.  
This is compared with other predictions that have been made using different
techniques. 
We highlight the significantly high bias for the H\/I distribution at small
scales.
This aspect has not been discussed before.  
We then discuss the prospects for detection with  the MWA, GMRT and
the hypothetical MWA5000.  
The MWA5000 can detect visibility correlations at large angular scales at all
redshifts in the post-reionization era. 
The GMRT can detect visibility correlations at lower redshifts, specifically
there is a strong case for a survey at $z \simeq 1.3$. 
We also discuss prospects for direct detection of rare peaks in the \h1
distribution using the GMRT. 
We show that direct detection should be possible with an integration time that
is comparable to, or even less than, the time required for a statistical
detection. 
Specifically, it is possible to make a statistical detection of the \h1
distribution by measuring the visibility correlation, and, direct detection of
rare peaks in the \h1 distribution at $z \simeq 1.3$ with the GMRT in less
than $1000$ hours of observations. 
\end{abstract}


\begin{keywords}
methods: N-Body simulations, cosmology: large scale structure of
the universe, galaxies: evolution, radio-lines: galaxies
\end{keywords}


\section{Introduction}

Large scale structures in the universe are believed to have formed by
gravitational amplification of 
small perturbations \citep{1980lssu.book.....P, 1989RvMP...61..185S,
1999coph.book.....P, 2002tagc.book.....P, 2002PhR...367....1B}.  
Much of the matter in galaxies and clusters of galaxies is the so
called dark matter that is believed to be weakly interacting and
non-relativistic \citep{1987ARA&A..25..425T, 2009ApJS..180..330K}.
Dark matter responds mainly to gravitational forces, and by virtue of
larger density than baryonic matter, assembly of matter into haloes
and large scale structure is driven by gravitational instability of
initial perturbations.  

Galaxies are believed to form when gas in highly over-dense haloes
cools and collapses to form stars in significant numbers
\citep{1953ApJ...118..513H, 1977MNRAS.179..541R, 1977ApJ...211..638S,
1977ApJ...215..483B}.  
The formation of first stars \citep{2007ARA&A..45..565M,
2007ARA&A..45..481Z, 2004ARA&A..42...79B} in turn leads to emission of
UV radiation that starts to ionize the inter-galactic medium (IGM).
The period of transition of the IGM from a completely neutral to a
completely ionized state is known as the epoch of reionization (EoR),
e.g., see \citet{2001ARA&A..39...19L}. 
Observations indicate that the process of reionization was completed before $z
\sim 6$ \citep{2006ARA&A..44..415F, 2001AJ....122.2850B,
  2006AJ....132..117F}. 
Many possible sources of ionizing radiation have been considered,
although stellar sources are believed to be the most plausible
candidates, see, e.g., \citet{2009MNRAS.397..971B}.

Prior to the EoR, almost all the Hydrogen in the universe is in atomic form. 
Through the EoR Hydrogen is ionized till we are left with almost no \h1 in the
inter-galactic medium (IGM) and almost all the \h1 resides in the inter stellar
medium (ISM) of galaxies.  
We focus on the post-reionization era in this work and our aim is to make
predictions about the distribution of \h1 in this regime. 
It had been proposed by \citet{1972A&A....20..189S, 1975MNRAS.171..375S} that
the hyperfine transition of Hydrogen may be used to probe primordial galaxies.  
This problem has been approached in past from the perspective of making
specific predictions for existing instruments like the Giant Meterwave Radio
Telescope (GMRT)\footnote{See http://gmrt.ncra.tifr.res.in/ 
  for further details.} 
\citep{1993MNRAS.265..101S, 1995MNRAS.272..544K, 1997MNRAS.289..671B,
  1999ASPC..156....9B, 2001JApA...22...21B, 2001JApA...22..293B,
  2003ASPC..289..251B, 2004JApA...25...67B, 2005MNRAS.356.1519B}, or upcoming
instruments like the the MWA\footnote{The Murchison Widefield Array (MWA).
  More details are available at http://www.mwatelescope.org/},
ASKAP\footnote{The Australian Square Kilometer Array 
  Pathfinder. See http://www.atnf.csiro.au/projects/askap/ for details.}
MeerKAT\footnote{The South African Square Kilometer Array Pathfinder.  See
  http://www.ska.ac.za/meerkat/overview.php for details.} and 
SKA\footnote{The Square Kilometer Array (SKA).  See
  http://www.skatelescope.org/ for details.}, 
e.g., see \citet{2008MNRAS.383.1195W}. 
Much work in the last decade has focused on making predictions for the power
spectrum of fluctuations in \h1, with an implicit assumption that it is easier
to make a statistical detection than a direct detection. 
Further, it has been argued that \h1 can be used as a tracer of the large
scale structure and observations can be used to constrain cosmological
parameters with a special emphasis on observations of the baryon acoustic
oscillations (BAO) in the power spectrum \citep{2008arXiv0812.0419V,
  2008PhRvL.100i1303C, 2009PhRvD..79h3538B}. 
Most of the upcoming instruments are sensitive to the power spectrum of
fluctuations at large scales, certainly larger than the scale of non-linearity
at the relevant epoch, and hence a significant fraction of the work done in
terms of making predictions is based on linear theory or approximations that
work well in the linear and quasi-linear regime. 
The halo model has also been used for predicting the spectrum of fluctuations
in the post-reionization universe \citep{2009arXiv0908.2854W,
  2009arXiv0912.2130W}.  

In this study we revisit the issue of predicting fluctuations and use high
resolution N-Body simulations. 
This allows us to study fluctuations at scales comparable to, or even smaller
than the scale of non-linearity. 
We use dark matter simulations along with simple ansatz for assigning neutral
Hydrogen in order to make predictions of fluctuations in surface brightness
temperature. 
Our model is discussed in detail in \S\ref{sec_h1model} where we 
summarize our knowledge of the \h1 distribution at high redshifts
and motivate the assignment schemes we use in this work .
The N-Body simulations used by us are described in \S\ref{subsec_nbody}.
Results of the HI signal are presented in \S\ref{sec_results}.
We then move on to review the relation between the source flux and observables
in radio interferometers, i.e. visibilities, 
in \S\ref{sec_sig_noise} and also discuss the sensitivity of interferometers.
Finally we look at the prospects of detection, both statistical and of 
rare peaks in \S\ref{sec_detection}.
We conclude with a discussion in \S{\ref{sec_discussion}}.

\section{Modeling the H\/I distribution}
\label{sec_h1model}

In this section we describe our model of the \h1 distribution at high
redshifts. 
Our knowledge of the \h1 distribution in the universe is derived mainly from
QSO absorption spectra, where the gas absorbs in the Lyman-$\alpha$ transition
of the Hydrogen atom.
We know from observations of these absorption spectra that much of
the inter-galactic medium (IGM) is highly ionized and does not contain
a significant amount of neutral Hydrogen.
Most of the neutral Hydrogen resides in relatively rare damped Lyman-$\alpha$
systems (DLAS) \citep{2005ARA&A..43..861W}. 
DLAS and other high column density absorption features are believed to arise
due to gas within galaxies \citep{2000ApJ...534..594H, 2001ApJ...559..131G}.
It is possible to make a quantitative estimate of the total neutral Hydrogen
content in DLAS and study the evolution of the total neutral Hydrogen content
of the universe \citep{1996MNRAS.283L..79S, 2000ApJS..130....1R,
  2005MNRAS.363..479P}.
These observations indicate that at $1 \leq z \leq 5$, the neutral Hydrogen
content of the universe is almost constant with a density parameter of
$\Omega_{\h1} \simeq 0.001$\footnote{These observational constraints are given
  as the ratio of the comoving density of neutral Hydrogen to the present day
  critical density.  We adopt the same convention here.}. 

At low redshifts, the \h1 content can be estimated more directly
through emission in the Hyperfine transition.     
Observations in the local universe indicate a much lower neutral Hydrogen
content than seen at $ z \geq 1$ \citep{2005MNRAS.359L..30Z}.
The neutral gas fraction in galaxies at intermediate redshifts appears to be
much higher than in galaxies in the local universe \citep{2007MNRAS.376.1357L,
  2009MNRAS.399.1447L}, i.e., the neutral gas fraction in galaxies appears to
increase rapidly with redshift $z$.

The spin temperature couples to the gas temperature through collisions of
atoms with other atoms, electrons, ions and also the Wouthuysen-Field
effect \citep{1956ApJ...124..542P, field1958, 1959ApJ...129..536F,
  1952Phy....18...75W, 1952AJ.....57R..31W, 2005ApJ...622.1356Z,
  2006PhR...433..181F, 2007MNRAS.379..130F, 2007MNRAS.374..547F}.
Observations of $21$~cm absorption by DLAS indicate that the spin temperature
is orders of magnitude higher than the temperature of the cosmic microwave
background radiation (CMBR) at corresponding redshifts
\citep{2000MNRAS.318..303C, 2009MNRAS.396..385K}.   
This implies that the emission in the $21$~cm hyperfine transition can be
safely assumed to be proportional to the density of neutral Hydrogen, see,
e.g., \citet{2006PhR...433..181F}. 
\begin{eqnarray}
\delta T_b(z) &=& 4.6~mK \left(1 - \frac{T_{cmb}}{T_s}\right) (1+z)^2
\frac{H_0}{H(z)}  \nonumber \\
&& ~~~~~ \times x_{\h1} \left(1 + \delta\right) \left[\frac{H(z)}{(1+z)
    (dv_\parallel/dr_\parallel)}\right] \nonumber \\ 
 &\simeq& 7.26mK (1+z)^2 \frac{H_0}{H(z)} \frac{M_{_{\h1}}}{10^{10}\msun} 
\left(\frac{L}{1Mpc}\right)^{-3} \nonumber \\
&& ~~~~~ \times \left[\frac{H(z)}{(1+z) (dv_\parallel/dr_\parallel)}\right]
\label{eqn_tb}
\end{eqnarray}
where $x_{\h1}$ is the fraction of Hydrogen in neutral form, $\delta$ is the
density contrast of the gas distribution, $T_{cmb}$ is the temperature of the
CMBR $T_s$ is the spin temperature defined using the relative occupation of
the two levels for the hyperfine transition: 
\begin{equation}
\frac{n_1}{n_0}=\frac{g_1}{g_0}\exp\left\{-\frac{T_\star}{T_s}\right\},
\label{eq_spin}
\end{equation}
where subscripts $1$ and $0$ correspond to the excited and ground
state levels of the hyperfine transition, $T_\star=h\nu/k_B=68$~mK is the
temperature corresponding to the transition energy, and
$(g_1/g_0)=3$ is the ratio of the spin degeneracy factors of the
levels.

Observations indicate that neutral gas is found only in galaxies in the post
reionization universe. 
We also know that at very low redshifts galaxies in groups and clusters do not
contain much neutral gas. 
As cold gas is associated with galaxies, we may assume that it exists only in
haloes that are more massive than the Jeans mass. 
Further, we may assume that neutral gas is predominantly found in galaxies and
not in larger haloes that may contain several large galaxies. 
Jeans mass for haloes in a photo-ionizing UV background depends on the shape of
the spectrum of the ionizing radiation, typically we expect gas in haloes with
a circular velocity in excess of $60$~km/s to cool, fragment and form stars. 
The mass within the virial radius is related to the circular velocity and the
collapse redshift as:
\begin{equation}
M_{vir} \simeq 10^{10} ~{\mathrm M}_\odot \left( \frac{v_{circ}}{60 {\rm
      km/s}}\right)^3 \left(\frac{1 + z_c}{4}\right)^{-3/2} 
\end{equation}
Simulation studies show that DLAS can reside in haloes with even lower
circular velocities \citep{2008MNRAS.390.1349P}, i.e., haloes with mass lower
than the expected Jeans mass can contain significant amount of neutral
Hydrogen.  
The gas in these haloes is able to self shield and maintain a significant
amount of \h1 even though the amount of gas is insufficient for sustaining
star formation\footnote{Haloes that collapse much earlier may have hosted star
formation, indeed the ISM of such galaxies may have been blown away due to
feedback.  However, few haloes with a low circular velocity that collapse very
early are likely to survive without merger well into the post-reionization
era.  Errors arising out of \h1 assignment to low circular velocity haloes
should not lead to significant changes in our calculations.  Errors arising
from this will generally lead to an under estimate of the signal (see
discussion of dependence of bias on the lower mass limit), hence our
predictions may be considered to be a lower limit on the signal.}. 
We use this input and impose a lower cutoff in circular velocity of
$30$~km/s, i.e., haloes with a lower circular velocity are not assigned any
\h1.  
This lower limit is appropriate for $z \sim 3$ and it is likely that the limit
shifts to higher values at lower redshifts.  
We keep the value fixed at $30$~km/s in order to make a conservative estimate
of clustering of such haloes. 
\citet{2008MNRAS.390.1349P} also found that haloes much more massive than a
few times $10^{11}$~M$_\odot$ do not host significant amount of \h1.  
This is consistent with observations in the local universe where galaxies in
groups and clusters of galaxies contain very little \h1. 
Some observations suggest that the neutral fraction in galaxies in outer parts
of clusters of galaxies at intermediate redshifts may be high compared to
corresponding galaxies in the local universe \citep{2009MNRAS.399.1447L}.
Assignment for \h1 in more massive haloes is done in such a way as to ensure
that very massive haloes have a zero or negligible fraction of the total gas
in neutral form.
In case larger haloes have more \h1 than we assume, our results under estimate
the true signal and may be considered as a lower bound.
The transition scale for the higher masses is chosen to coincide with
$200$~km/s. 
Given that it is easier to estimate mass than circular velocity for haloes
identified in simulations, in particular for haloes with a small number of
particles, we choose to translate the threshold in circular velocity to a
threshold in mass assuming the collapse redshift and the redshift at which the
system is being observed to be the same, i.e., $z_c = z$.  
While this may lead to inaccuracies regarding inclusion of haloes close to the
low mass end, we do not expect it to influence the results in a significant
manner. 
The minimum and the maximum masses for the redshifts considered here are
listed in Table~\ref{table_h1_assign}.

\begin{table}
\caption{The table lists the minimum mass of haloes $\mmin$ that hosts \h1 in
  the simulated \h1 distribution.  The first column lists the redshift where
  simulated maps are constructed, the second column lists $\mmin$ we get by
  assuming a lower bound of $30$~km/s on the circular velocity of haloes, and
  column 4 lists the actual minimum halo mass used in construction of the
  simulated maps.  All masses are given in units of solar mass
  M$_\odot$.  Column 3 lists the characteristic maximum mass of haloes
  computed using an upper bound on circular velocity of $200$~km/s and Column
  5 shows the actual upper bound used while making simulated maps.  Column 6
  lists the \h1 fraction of the total baryonic mass if we use the first \h1
  assignment scheme (see Eqn.~\ref{eq_fh1_sharp}.}
\label{table_h1_assign}
\begin{center}
\begin{tabular}{||c|c|c|c|c|c||}
\hline
\hline
$z$ & $M_{min}$ & $M_{max}$ & $M_{min}^{sim}$& $M_{max}^{sim}$& $F_1
\Omega_{nr} / \Omega_{b}$ \\ 
\hline
\hline
5.1 &$10^{8.83}$&$10^{11.30}$&$10^{8.81}$&$10^{11.30}$&$0.24$\\
\hline
3.4 &$10^{9.04}$&$10^{11.52}$&$10^{9.02}$&$10^{11.50}$&$0.15$\\
\hline
1.3 &$10^{9.43}$&$10^{11.92}$&$10^{9.43}$&$10^{11.90}$&$0.11$\\
\hline
\hline
\end{tabular}
\end{center}
\end{table}

Given a halo of mass $M$, all particles in it are assigned an equal \h1 mass
which is a fraction of their total mass.  
We describe three kinds of mass assignments here.
\baq
F_1(M) &=&  f_1 ~~~~~~~~~~~~~~~~~~~~~~~~~~~~~~~ (\mmin \leq M \leq
\mmax) \label{eq_fh1_sharp} \\
F_2(M) &=& 
\frac{f_2}{1+\left(\frac{M}{\mmax}\right)^2} ~~~~~~~~~  (\mmin \leq
M)  \label{eq_fh1_smooth1} \\ 
F_3(M) &=& 
\frac{f_3}{1+\left(\frac{M}{\mmax}\right)} ~~~~~~~~~~~  (\mmin \leq M
) \label{eq_fh1_smooth2}  
\eaq
Here $F_k(M)$ is the mass fraction of \h1 in a given halo.  
The fraction of baryons that is in the form of neutral Hydrogen is given by
$F_k(M) \Omega_{nr} / \Omega_{b}$, where $\Omega_b$ is the density parameter of
Baryons and $\Omega_{nr}$ is the density parameter for non-relativistic
matter.  
For the best fit model for WMAP-5 \citep{2009ApJS..180..330K} the ratio has 
the value $\Omega_{nr}/\Omega_b \simeq 1/0.17 \simeq 6$.

In all three cases the constant $f_k$ is determined by normalizing the \h1
mass in the simulation volume to $\omegah1= 0.001$, the density parameter of
\h1 indicated by observations at relevant redshifts
\citep{2005MNRAS.363..479P}.  
It is noteworthy that this normalization requires a significant fraction
of gas in haloes in the mass range between $\mmin$ and $\mmax$ be neutral at
high redshifts. 
In the assignment scheme with the sharp cutoff (scheme~1,
Eqn.~\ref{eq_fh1_sharp}), the \h1 fraction is the same in all haloes with mass
between $\mmin$ and $\mmax$. 
In the second assignment scheme the \h1 fraction in a halo
decreases monotonically with increasing halo mass, and for large halo masses
the \h1 mass goes to zero.
This scheme applies to the physical situation where very large haloes do not
have any neutral Hydrogen.
In scheme 3, the \h1 fraction decreases monotonically with increasing halo
mass but the \h1 mass goes to a constant value.
The motivation of this scheme is to allow for some neutral Hydrogen surviving
in massive haloes in galaxies.
In the following section, we compare the resulting distribution of \h1 with
the three assignment schemes described above. 
The present work is the first one where mass resolution of simulations is
adequate for resolving the smallest haloes that may host \h1 and the
simulation volume is also sufficient to make statements about the large scale
distribution. 
We also ensure that the finite size of the simulation volume does not affect
the results presented here. 

\begin{table}
\caption{Columns 1 and 2 list the size of the box and the number of
 particles used in the simulations. Columns 3 and 4 give the mass and force 
 resolution of the simulations, while columns 5 and 6 tell us the redshift at
 which  the simulations were terminated and the redshifts for which the
 analysis were done}
\begin{center}
\begin{tabular}{||l|l|l|l|l|l||}
\hline\hline
$L_{box}$
& $N_{part}$
& $m_{part}$
& $\epsilon$
& $z_f$ 
& $z_{out}$\\
$(h^{-1}Mpc)$
& $$
& $(h^{-1}\msun)$
& $(h^{-1}kpc)$
& $$ 
& $$\\
\hline\hline
$23.04$
& $512^3$
& $6.7 \times 10^6$
& 1.35
& 5.0
& 5.04 \\
\hline
$51.20$
& $512^3$
& $7 \times 10^7$
& 3.00
& 3.0
& 3.34 \\
\hline
$76.80$
& $512^3$
& $2.3 \times 10^8$
& 4.50
& 1.0
& 1.33 \\
\hline\hline
\label{table_nbody_runs}
\end{tabular}
\end{center}
\end{table}

\subsection{$N$-Body Simulations}
\label{subsec_nbody}

We use gravity only simulations run with the TreePM code
\citep{2002JApA...23..185B, 2003NewA....8..665B, 2009RAA.....9..861K}. 
The suite of simulations used here is described in the
Table~\ref{table_nbody_runs}. 
The cosmological model and the power spectrum of fluctuations corresponds to
the best fit model for WMAP-5:
$\Omega_{nr} = 0.26$, $\Omega_{\Lambda} =0.74 $, $n_s = 0.96$, 
$\sigma_8 = 0.79$, $h=0.72$, $\Omega_b h^2 = 0.02273$
\citep{2009ApJS..180..330K}. 

We use the Friends-of-Friends (FOF) \citep{1985ApJ...292..371D}
algorithm with a linking length $l=0.2$ to identify haloes and construct a
halo catalog. 
The \h1 assignment schemes are then used to obtain the distribution of neutral
Hydrogen. 

Several existing and upcoming instruments can probe the post-reionization
universe using redshifted \h1 emission. 
The GMRT can observe redshifted \h1 emission from a few
selected redshift windows whereas most other instruments have continuous
coverage in redshift bounded on two sides.
We choose to focus on the GMRT windows, as these are representative of the
range of redshifts in the post-reionization universe. 
In particular we will focus on the following redshift windows of GMRT:
$z_{out}=5.04$, $3.34$, $1.33$. 
Finite box effects can lead to significant errors in the distribution of
haloes that host galaxies, apart from errors in the abundance of haloes of
different masses \citep{2006MNRAS.370..993B, 2009MNRAS.395..918B}. 
The choice of simulations used in this work ensures that such effects do not
contribute significantly.  
Previous studies have indicated that at $z \simeq 0$, we need a simulation box
with $L_{box} \geq 140 $~h$^{-1}$Mpc for the finite size effects to be
negligible \citep{2005MNRAS.358.1076B, 2006MNRAS.370..993B}.
On the other hand the mass resolution of particles decreases as the cube
of simulation volume. 
We balance the requirements of high mass resolution and a sufficiently large
box size by using different simulations for studying the \h1 distribution at
different redshifts. 
Details of the simulations are given in the table~\ref{table_nbody_runs}.

\begin{figure}
\includegraphics[width=3.2truein]{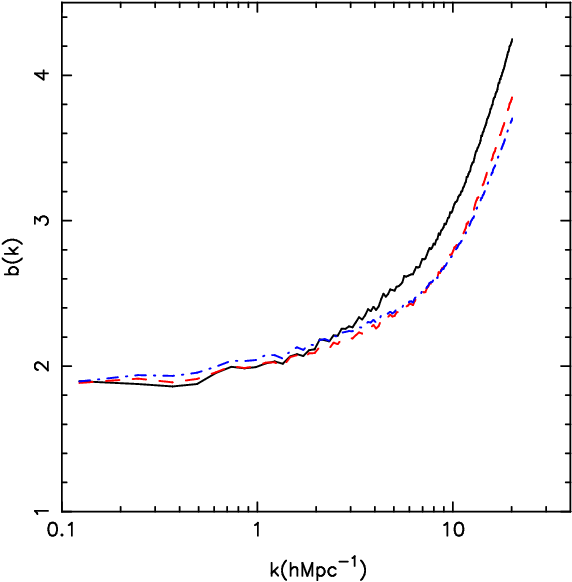} 
\caption{Effect of \h1 mass assignment type on bias. 
Solid,dashed and dot-dashed lines are for \h1 assignment types 
$1$, $2$ and $3$ as described in eqs.~\ref{eq_fh1_sharp}-\ref{eq_fh1_smooth2}.  
Bias is computed at $z=3.34$ for the $51.2$h$^{-1}$Mpc box with
$\mmax=10^{11.5}\msun$. 
 \label{fig_bias_fix}. }
\end{figure}
\begin{figure}
\includegraphics[width=3.2truein]{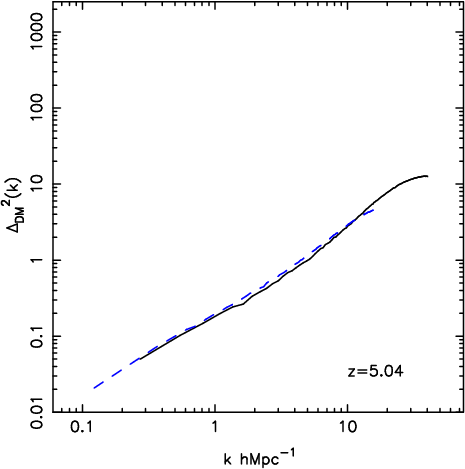} 
\includegraphics[width=3.2truein]{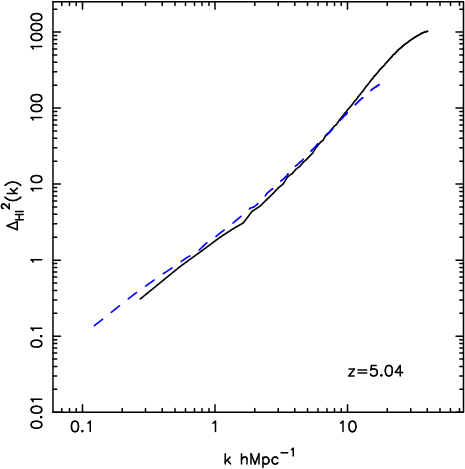}
\caption{\emph{Top}: The dark matter power spectrum in the simulations with
  $L_{box} = 23$~h$^{-1}$Mpc and $L_{box} = 51.2$~h$^{-1}$Mpc at $z=5.1$.  The
  agreement in the two curves indicates that the finite box size effects do
  not lead to errors in the smaller box at the redshift of interest. 
\emph{Bottom}: \h1 power spectrum in the same simulations, see text for
details.  Show the effect of finite box size on the simulated \h1 power
spectrum, good agreement between the two curves imples that the error is
small.} 
\label{fig_boxcheck}
\end{figure}
\begin{figure}
\includegraphics[width=3.2truein]{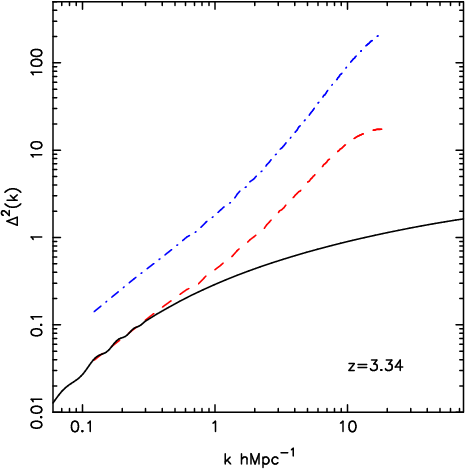}
\caption{The power spectrum of fluctuations: the solid line shows the linearly
  extrapolated power spectrum, the dashed line shows the non-linear dark matter
  power spectrum and the dot-dashed line shows the \h1 power spectrum.  All
  power spectra are for $z=3.34$.  The dark matter and the \h1 power spectra
  have been computed with the simulation with  $L_{box} = 51.2$~h$^{-1}$Mpc.}
\label{fig_nonlin}
\end{figure}
\begin{figure}
\includegraphics[width=3.2truein]{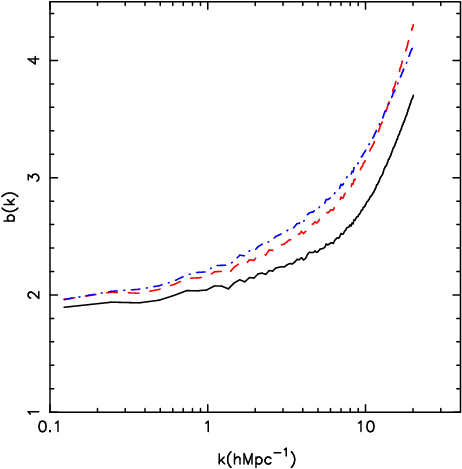}
\caption{The dependance of bias on $\mmin$ and $\mmax$.  
The solid curve shows $b(k)$ for the values of $\mmin$ and $\mmax$ shown in
Table~\ref{table_h1_assign}. 
The dashed line shows $b(k)$ when we use the reference value for $\mmax$
but increase $\mmin$ to twice the reference value. 
The dot-dashed curve shows $b(k)$ when $\mmax$ is chosen to be higher:
$10^{11.9}$~M$_\odot$, while $\mmin$ is kept at the reference alue.
We see that as the characteristic mass of haloes with \h1 increases, $b(k)$
increases. 
All curves are for $z=3.34$ and have been computed with the simulation with
$L_{box} = 51.2$~h$^{-1}$Mpc.} 
\label{fig_bias_mmax}
\end{figure}
\begin{figure}
\includegraphics[width=3.2truein]{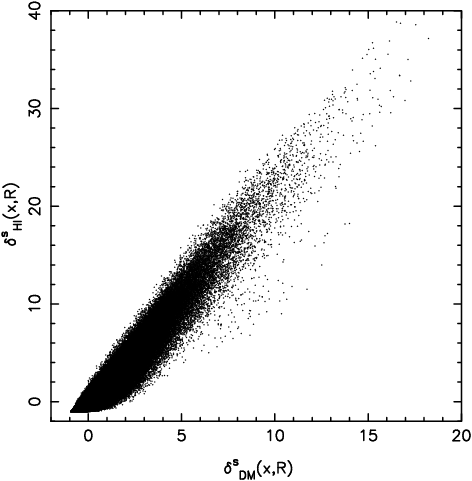}
\caption{This figure shows a scatter plot of $\delta_{\h1}$ smoothed at a
  scale of $3 h^{-1}$Mpc with spherical top hat window, 
  plotted as a function of $\delta_{DM}$ smoothed at the same
  scale.  The figure shows a random subset of points from the simulation with
  $L_{box} = 51.2$~h$^{-1}$Mpc at $z=3.34$.}
\label{fig_stoch}
\end{figure}

\section{Results}
\label{sec_results}

Given an \h1 mass assigment for particles in a simulation, we can proceed to
compute the expected signal from the \h1 distribution by making mock radio
maps and spectra.
We also compute the power spectrum in both the real and the redshift space. 
For redshift space calculations, we use the peculiar velocity information of
particles in haloes. 
Unlike earlier studies, we resolve haloes of individual galaxies and hence the
internal velocity dispersion is naturally accounted for and there is no need
to add it by hand \citep{1995MNRAS.272..544K, 1997MNRAS.289..671B,
  2003ASPC..289..251B}. 

We use the clouds-in-cell (CIC) smoothing to interpolate densities from
particle  locations to grid points for the purpose of computing densities and
the power spectrum. 
It is convenient to express the \h1 power spectrum in terms of the brightness
temperature, though in comparisons with the dark matter power spectrum we
revert to the usual dimensionless form.

We define the real and redshift space scale dependant bias by the 
ratio of the corresponding dark matter and \h1 power spectra.
\baq
b(k) &=& \left[\frac{\ph1k}{P_{_{DM}}(k)}\right]^{1/2}\label{eq_bias_real}\\
b^s(k) &=& \left[\frac{P^s_{_{\h1}}(k)}{P^s_{_{DM}}(k)}\right]^{1/2}
\label{eq_bias_red}
\eaq
We start by checking the effect of \h1 assignment scheme on the resulting
distribution. 
We have computed the bias $b(k)$ at $z=3.34$ for the $L_{box} =
51.2$h$^{-1}$Mpc simulation using the three schemes introduced above.
The results are shown in Figure~\ref{fig_bias_fix} for
the three assignment types described in
Eqs.~(\ref{eq_fh1_sharp}-\ref{eq_fh1_smooth2}).  
We see that at large scales the three assignment schemes give very similar
results, whereas there is some disagreement between the assigment scheme
Eqn.(\ref{eq_fh1_sharp}) and the other two.  
The difference can be attributed to the fact that the scheme described in
Eqn.(\ref{eq_fh1_sharp}) puts more \h1 mass in haloes with masses near $\mmax$. 
The differences between the \h1 assignment schemes are relatively minor. 
We choose to work with the scheme described in Eqn.(\ref{eq_fh1_smooth2}) due
to a better physical justification, as discussed in \S\ref{sec_h1model}.
As mentioned before, if large mass haloes happen to have a larger \h1 fraction
than we assume then the signal will be larger than what we predict here.

It is noteworthy that while the bias is strongly scale dependent at large $k$
(small scales), it flattens out to a constant value at small $k$ (large
scales). 

In order to check for the effects of a finite box size, we carried out a test
for the $23$~h$^{-1}$Mpc simulation box. 
Instead of using the range of halo masses for \h1 assignment that is
appropriate for $z =5.1$, we work with a slightly smaller range so that haloes
of these masses can be found in the $51.2$~h$^{-1}$Mpc simulation as well. 
Figure~\ref{fig_boxcheck} shows the dark matter power spectrum (Top panel)
and the \h1 power spectrum (Lower panel) with the \h1 assignment restricted to
a smaller range of masses, as described above. 
We see that the dark matter power spectra from the two simulations agree
through the range of scales where there is an overlap. 
The \h1 power spectra also agree, though not as well as the dark matter power
spectra. 
These differences are so small that we do not expect these to affect the final
results in a significant manner. 
The difference can be attributed to the fact that clustering of haloes is
affected more strongly by the box size effects in simulations
\citep{1994ApJ...436..491G, 2005MNRAS.358.1076B}.

These tests validate our approach for assignment of \h1 to haloes, and also
show that the effects of a finite box-size are not significant at the level of
the power spectrum or the mass function.

\subsection{Bias}
\label{subsec_bias}

This is amongst the first studies of the \h1 distribution at high redshifts
where we resolve the smallest haloes that can host significant amount of \h1
while ensuring that the finite box size effects do not lead to an erronous
distribution of haloes.  
One of the points that we can address here is the effect of non-linear
clustering on the \h1 distribution and the scale dependance of bias. 
Some of these effects are illustrated in Figure~\ref{fig_nonlin} where we have
plotted the power spectrum of fluctuations: the solid line shows the linearly
extrapolated power spectrum, the dashed line shows the non-linear dark matter
power spectrum and the dot-dashed line shows the \h1 power spectrum.  
All power spectra are for $z=3.34$.  
The dark matter and the \h1 power spectra have been computed with the
simulation with $L_{box} = 51.2$~h$^{-1}$Mpc.  

It is apparent that at $k > 0.5$~h~Mpc$^{-1}$ the effects of non-linear
clustering significantly enhance the dark matter power spectrum.
At $k \sim 10$~h~Mpc$^{-1}$, the enhancement is close to an order of magnitude. 
This is of utmost interest for upcoming instruments that can resolve small
angular scales.

We find that bias $b(k)$ for the \h1 distribution is much greater than unity
at high redshifts. 
This is to be expected of galaxies at high redshifts
\citep{1996ApJ...461L..65F, 1996MNRAS.282..347M, 1998MNRAS.297..251B,
  1998MNRAS.299..417B, 1999MNRAS.304..175M, 1999MNRAS.305L..21B,
  2000MNRAS.314..546M, 2000MNRAS.311..793B, 2000ASPC..200...24R,
  2001MNRAS.323....1S, 2009arXiv0912.2130W}.   
We note that the bias is scale dependent, and leads to a larger enhancement
in the power spectrum at very small scales.  

The value of bias depends strongly on the choice of the characteristic mass of
haloes with \h1.
This is shown in Figure~\ref{fig_bias_mmax}.
All curves are for $z=3.34$ and have been computed with the simulation with
$L_{box} = 51.2$~h$^{-1}$Mpc. 
This figure illustrates that the bias at all scales varies monotonically
with the charactersitic mass of haloes with \h1. 
Variation is gentle at large scales but fairly strong at small scales. 
Therefore it is extremely important to have an accurate estimate of the
charactersitic masses of such haloes. 
Indeed, it has been pointed out that observations of the amplitude of
clustering in the \h1 distribution can be used to constrain masses of haloes
that host DLAS \citep{2008arXiv0804.1624W}.

While the preceeding figures describe the statistical bias computed from the
ratio of power spectra, Figure~\ref{fig_stoch} shows the stochasticity of bias
\citep{1999ApJ...520...24D} in the \h1 distribution.
This figure shows a scatter plot of $\delta_{\h1}$ smoothed at a scale of
$3$~h$^{-1}$Mpc with a spherical top hat window, plotted as a function of
$\delta_{DM}$ smoothed at the same scale.  
The figure shows a random subset of points from the simulation with  $L_{box}
= 51.2$~h$^{-1}$Mpc at $z=3.34$. 
The scatter about the average trend in the $\delta_{\h1}-\delta_{DM}$ is
significant, and increases as we go towards large overdensities in dark
matter.  
The scatter becomes small if we smooth the density distribution at much larger
scales.  

Similar results concerning the effect of non-linear gravitational clustering
and bias are obtained for other redshifts.
We concentrate on the evolution of bias and the power spectrum rather than go
into a very detailed discussion of the \h1 distribution at each redshift.
We discuss simulated maps in the following subsection.

\begin{figure*}
\begin{center}
\begin{tabular}{cc}
\includegraphics[width=3.2truein]{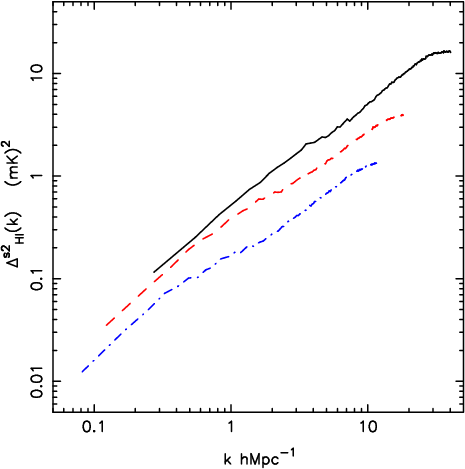} &
\includegraphics[width=3.2truein]{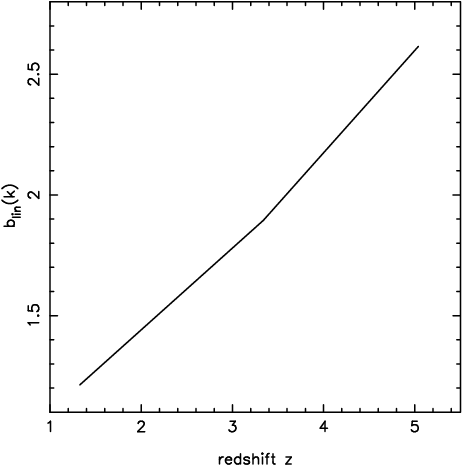}
\end{tabular}
\end{center}
\caption{\emph{Left}: Evolution of the \h1 powerspectrum $\dh1k$. 
Solid, dashed, dot-dashed lines are for redshifts $z=5.04$, $3.34$, $1.33$
respectively. \emph{Right}: Evolution of the \h1 linear bias.} 
\label{fig_evolution}
\end{figure*}

Figure~\ref{fig_evolution} (left panel) shows the evolution of the redshift
space power spectrum for the \h1 distribution. 
Curves show the $\Delta^2_{\h1}(k)$ as a function of $k$ for $z=5.1$ (solid
line), $z=3.34$ (dashed line) and $z=1.3$ (dot-dashed line). 
Unlike the power spectrum in real space, the enhancement at very small scales
is less strong and this is due to velocity dispersion within haloes at smaller
scales. 
Power at larger scales is enhanced due to the Kaiser
effect \citep{1987MNRAS.227....1K}, leading to a relatively stronger
enhancement at larger scales, hence the difference in shapes of the \h1 power
spectrum in the real space and the redshift space. 
Even though the amplitude of density perturbations increases with time, we see
that the amplitude of brightness temperature fluctuations decreases instead.
This is a manifestation of the decreasing correlation function for galaxies,
see, e.g., \citet{1998MNRAS.299..417B, 1998wfsc.conf..414R}.

The evolution of bias at large scales, denoted as $b_{lin}(k)$ to emphasise
that it refers to scales where clustering is still in the linear regime, is
shown in the right panel of Figure~\ref{fig_evolution}. 
We see that this decreases from around $2.5$ at $z=5.1$ to $1.2$ at $z=1.3$,
the variation is slightly steeper than $(1+z)^{1/2}$ and hence the gradual
lowering of the amplitude of the brightness temperature power spectrum.

\begin{figure*}
\begin{center}
\begin{tabular}{cc}
\includegraphics[width=3.2truein]{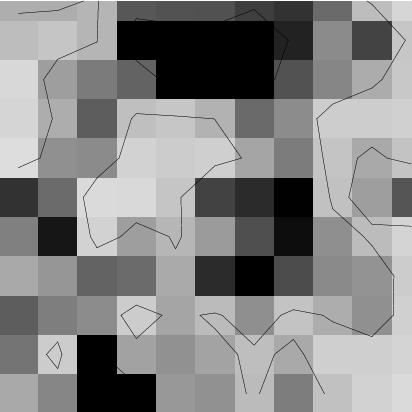} &
\includegraphics[width=3.2truein]{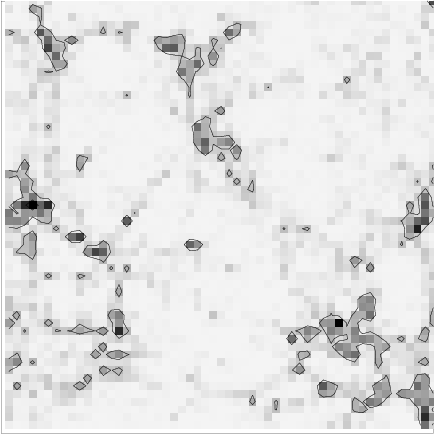} \\
\includegraphics[width=3.2truein]{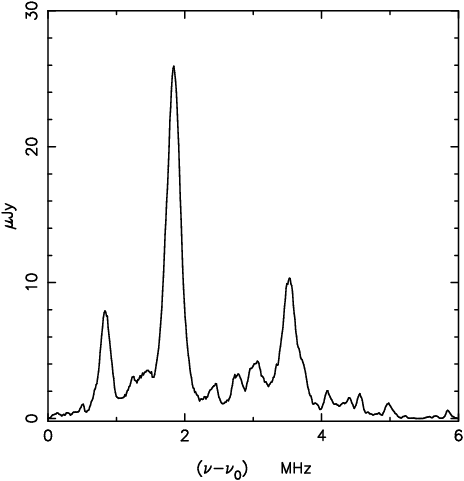} &
\includegraphics[width=3.2truein]{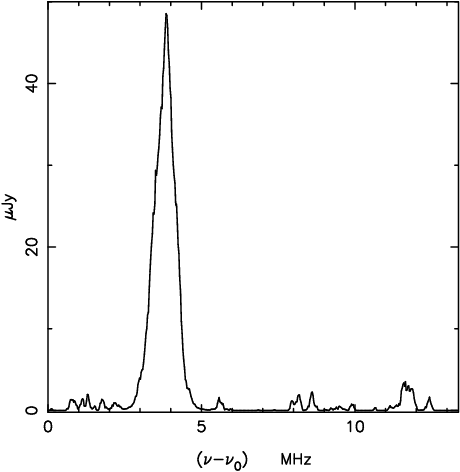} \\
\end{tabular}
\end{center}
\caption{The top row shows mock radio maps, the top-left panel is for $z=3.34$
  and the top-right panel is for $z=1.3$.  The pixel size is chosen to be
  the resolution of the central square of the GMRT.  
  The bandwidth of the map is $0.5$MHz and $1.0$MHz for $z=3.34$, $1.33$ 
  respectively. 
  The  corresponding contours mark regions with signal  
  ($9.2$, $5.75$ ,$1.15$)~$\mu$Jy and ($16.8$, $10.5$, $2.1$)~$\mu$Jy for
  $z=3.34$ and $1.33$ respectively.
  The mock maps show the brightest regions from the simulation.  As we can
  see, these regions have a large angular extent this fact may be used to
  enhance prospects of detection.  The lower panel shows spectra from the
  mock radio maps.  
  These spectra have been plotted for regions that are the
  size of the resolution of the central square of the GMRT for $z=3.34$ and 
  twice the resolution of the central square of GMRT for $z=1.33$.  
  We chose to plot the spectrum  through the brightest region in the
  simulation.  The lower-left panel corresponds to $z=3.34$ and the lower
  right panel shows the spectrum for 
  $z=1.3$.  The strength of the signal for $z=3.34$ is comparable to what we
  found in an earlier study.  The strentgh of the predicted signal at $z=1.3$
  is or the order of $0.05$~mJy, with an FWHM of around $1$~MHz.  The best {\it
    rms} sensitivity achieved in this band at the GMRT is $0.02$~mJy.  This
  has been achieved with a bandwidth $\simeq 32$~MHz.}
\label{fig_mockmaps}
\end{figure*}

\subsection{Radio Maps}
\label{subsec_radiomaps}

Figure~\ref{fig_mockmaps} shows the simulated radio maps and spectra. 
The usual conversion from \h1 density to signal has been used for this
\citep{2006PhR...433..181F}.  
The top row shows simulated radio maps, the top-left panel is for $z=3.34$
and the top-right panel is for $z=1.3$.  
The pixel size is chosen to be the same as the resolution of the central
square of the GMRT. 
The bandwidth of the map is $0.5$MHz and $1.0$MHz for $z=3.34$, $1.33$ 
respectively. 
The  corresponding contours mark regions with signal ($9.2$, $5.75$,
$1.15$)~$\mu$Jy for $z=3.34$ and ($16.8$, $10.5$, $2.1$)~$\mu$Jy for
$z=1.33$. 
As we can see in these maps, bright regions have a large angular extent.
This fact may be used to enhance prospects of detection \S\ref{sec_detection}. 

The lower panel shows spectra from the mock radio maps.  
These spectra have been plotted for regions that are the size of the
resolution of the central square of the GMRT.   
We have chosen to plot the spectrum through the brightest region in the
simulation.  
The lower-left panel corresponds to $z=3.34$ and the lower right panel shows
the spectrum for $z=1.3$. 
The peak flux is around $0.025$~mJy and the FWHM of the line is close to
$300$~kHz.  
The strength of the signal for $z=3.34$ is comparable to what we
found in an earlier study, although the line width is much smaller. 
The reason for a smaller line width is likely to be the supression of the \h1
fraction in very massive haloes.  Considering the
  peak for which 
the spectrum has been plotted here, it may be possible to make a $2\,
\sigma$  detection with about $2\times 10^3$ hours of observations with
the central square of the GMRT (\S\ref{sec_detection}). 
Given that the GMRT observes a much larger volume at these redshifts that the
simulation volume, it is highly likely that even rarer peaks in the \h1
distribution will be observed in a generic pointing
\citep{1993MNRAS.265..101S}. 

The strentgh of the predicted signal at $z=1.3$ is of the order of $0.05$~mJy,
with an FWHM of around $1$~MHz.  

Detection of rare peaks in the \h1 distribution is an exciting possibility. 
The size of the region represented in these rare peaks is fairly large, and it
should be possible to establish the total mass contained in these regions
using observations in other wavebands. 
This will allow us to estimate $\Omega_{\h1}$ in emission, and hence provide
an independent measurement of the cold gas fraction.

The time required for detection of \h1 at high redshifts with the GMRT
increases rapidly as we go to higher redshifts. 
This trend continues as we move to $z\simeq 5.1$ and therefore we do not
discuss those results in detail here. 

Instruments like the GMRT, MWA can in principle detect signal from the \h1 
distribution at high redshifts.
The angular resolution of the MWA is fairly poor and hence the effects of
non-linear clustering and scale dependant bias do not make a significant
impact on predictions. 
We shall discuss the prospects for detection with the GMRT and the MWA in
detail in the \S\ref{sec_detection}.

\begin{table*}
\caption{Characteristics of the GMRT, MWA and MWA 5000 that have been used for
  calculation of the visibility correlation and noise in the visibility correlation as well as noise in image
  are listed here.  The values of the system parameters for GMRT and MWA 
  were taken from \emph{http://gmrt.ncra.tifr.res.in/} and
  \citet{bowman-thesis} respectively.
}
\label{table_system_param}
\begin{tabular}{cccccccc}
\hline
Instrument & Frequency (MHz) &  $\Delta\nu$ (MHz)&  B (MHz) &$\Delta
 U$ & $\theta_0$ (degree )& $T_{sys}$ (K)&$A_{eff} (m^2)$ \\
\hline
\hline
GMRT & 610 & 1 & 16 &32 & 0.56 & 102& 1590\\
\hline
&330&1& 16& 17&1.03& 106 & 1590\\
\hline
&235&1&16&12&1.45& 237&  1590\\
\hline
MWA & $<$ 300 & 1 & 32& 2 & 9 & 68 & 5.6\\
\hline
MWA5000 & $<$ 235&1 & 32 & 1.6 & 11.49 & 92 & 15 \\
\hline
\end{tabular}
\end{table*}

\section{Signal and Noise in Interferometers}
\label{sec_sig_noise}

In this section we briefly review the relation between the flux from sources
and the observable quantities for radio interferometers. 
We then proceed to a discussion of the sensitivity of interferometers and the
corresponding limitations arising from that. 

\subsection{Visibility Correlation}
\label{subsec_viscorr_int}

The quantity measured in radio-interferometric observations is the
visibility $V(\U,\nu)$ which is measured in a number of frequency
channels $\nu ~ - ~ \left(\nu + \Delta\nu\right)$ across a frequency
bandwidth $B$ for every pair of antennas in the array. 
The visibility is related to the sky specific intensity pattern $I_{\nu}(\Th)$
as  
\be
V(\U,\nu)=\int d^2 \Th A(\Th) I_{\nu}(\Th)
e^{ 2\pi \imath \Th \cdot \U}.
\label{eq_vis}
\e
The baseline $\U$ is ${ {\mathbf d}}/\lambda$  where ${{\mathbf d}}$
is the antenna separation projected in the plane perpendicular to the line of
sight.  
$\Th$ is a two dimensional vector in the plane of the sky with
origin at the center of the field of view, and $A(\Th)$ is the 
beam  pattern of the individual antenna. 
For the GMRT this can be well approximated by Gaussian
$A(\Th)=e^{-{\theta}^2/{\theta_0}^2}$, where $\theta = \left|\Th\right|$.   
We use $\theta_0=0.56^{\circ}$ at $610$~MHz for the GMRT, and it
scales as the inverse of frequency.  

We next consider the visibility-visibility correlation (hereafter only
visibility correlation) measured at two different baselines $\U$ and
$\U+{{\mathbf \Delta \mathbf U}}$ and at two frequencies $\nu$ and $\nu+\Delta
\nu$. 
As argued in \cite{2005MNRAS.356.1519B} the visibilities at baselines
$\U$ and $\U+{{\mathbf \Delta \mathbf U}}$ will be correlated only if
$|\mathbf\Delta \U|<1/\pi\theta_0$. 
Visibilities at different frequencies are expected to be correlated only if
the flux from sources is correlated.
This is certainly true of continuum radiation. 
In case of spectral lines, like the redshifted HI $21$~cm line being considered
here, visibilities are expected to be correlated over a range of frequencies
comparable to width of spectral lines.  
For generic applications, we can take this to be about $1$~MHz though we can
use the simulated radio maps for a more refined model. 
As a first approximation, the visibilities at frequencies $\nu$ and
$\nu+\Delta \nu$ can be assumed to be uncorrelated for $\Delta \nu>1$~MHz
\citep{2007MNRAS.378..119D} for range of baselines of our interest. 
We use the power spectra to compute visibility correlations 
using the relation enunciated by \citet{2001JApA...22..293B}.
To calculate the strength and nature of visibility correlation of the
HI fluctuations we
consider the visibility correlation at same baselines and frequencies which
can be written as  (for details see \citet{2005MNRAS.356.1519B})
\be
V V^*(\U,\, \nu) \equiv \langle V(\U,\, \nu)V^*(\U,\, \nu)\rangle=\frac{{\bar
    I}^2_{\nu}\theta_0^2}{2r_{\nu}^2} \int_0^{\infty} dk_{\parallel} P_{\rm
  HI}(k) 
\label{eq_vis_cor}
\e 
where $ {\bar I}_{\nu}$ is mean HI specific intensity, $r_{\nu}$ is
comoving distance corresponds to frequency 
$\nu=1420/\left(1+z\right)$. ${\mathbf k}=k_{\parallel}{{\mathbf m}}+{2\pi
  \U}/{r_{\nu}}$, where ${ {\mathbf m}}$ is unit vector along the line of
sight.  
Here the quantity $\bar I$ is proportional to $\Omega_{\h1}$, hence the
visibility correlation scales as the square of $\Omega_{\h1}$.

We next shift  our focus on the redshift space HI power spectrum $P_{\h1}(k,
\, z)$ \citep{2004MNRAS.352..142B}. 
We calculate $P_{\h1}(k, \, z)$ from N-Body simulations and put this in
Eqn.~\ref{eq_vis_cor} to calculate the visibility correlation.
As simulations have a finite box size, we patch it with a linearly
extrapolated power spectrum with a constant bias at small $k$. 
This is especially appropriate for high redshifts where the field of view of
the GMRT is larger than the simulation box. 
MWA, of course, has a much larger field of view and the statement applies to
that as well.

At sufficiently large scales one can write $P_{\h1}(k, \, z)$ as
\citep{1987MNRAS.227....1K} 
\be
P_{\h1}(k, \, z)= b^2
\left[1+\beta(k)\frac{k^2_{\parallel}}{k^2}\right]^2 P(k,z).
\label{eq_pk_h1}
\e
Here $b$ is the scale independent bias at large scales and 
$\beta(k)=f/b$ where $f=\Omega_{nr}^{0.6}$ is the redshift distortion
parameter. 
$P(k,\, z) $ is matter power spectrum at redshift $z$ and is dominated by the
spatial distribution of dark matter. 

\begin{figure}
\begin{center}
\includegraphics[width=2.truein,angle=270]{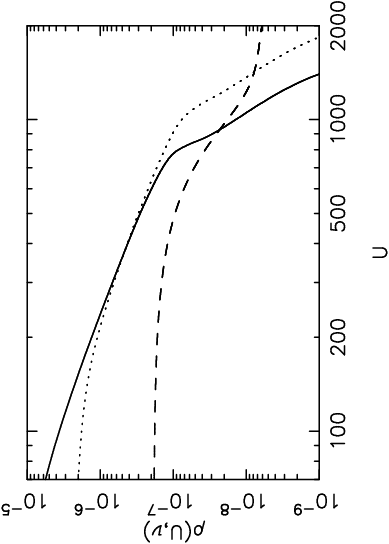} 
\end{center}
\caption{This shows the normalized baseline distribution function
  $\rho(U,\nu)$ as a function of baseline $U$ for three instruments
  (GMRT, MWA and MWA5000) for frequency $\nu=330\, {\rm MHz}$.}
\label{rho}
\end{figure}

\subsection{System Noise in the Visibility Correlation}
\label{subsec_sysnoise}

The noise {\it rms}. in real part in each visibility for single polarization
is 
\be
V_{rms}=\f{\sqrt{2}k_BT_{sys}}{A_{eff}\sqrt{\Delta \nu \Delta t}}
\e
where $T_{sys}$ is the total system temperature, $k_B$ is the Boltzmann
constant, $A_{eff}$ is the effective collecting  area of each
antenna, $\Delta \nu$ is the channel width and $\Delta t$ is correlator
integration time. 
This can be written in terms of the antenna sensitivity
$K={A_{eff}}/{2\,k_B}$ as  
\be
V_{rms}=\frac{T_{sys}}{K\sqrt{2 \Delta \nu \Delta t}}
\e
The noise variance in the visibility correlation is written as \citep{ali08}  
\be
\sigma^2_{VV}=\frac{8\,V_{rms}^4}{N_p}
\e
where $N_p$ is the number of visibility pairs in a particular $U$ bin and
frequency band $B$. 
The \h1 signal is correlated but the system noise is taken to be uncorrelated
in visibilities. 

We next consider how $N_p$ is calculated. 
We consider an elementary grid with cell of area $\Delta U^2$ in the $u-v$
plane where the signal remains correlated. 
Total observation time we consider is $T$. 
We further define $\rho(U,\nu)$ (circularly symmetric) to be the  baseline 
fraction per unit area per in the baseline range $U$ and $U+\Delta U$
and is normalised:  $\int d^2U \rho(U,\nu)=1$. 
Number of visibility pairs in a given $U$, $\Delta\nu$ bin in the grid is 
\be
\frac{1}{2}\left[ \frac{N(N-1)}{2}\frac{T}{\Delta t}\Delta U^2
  \rho(U,\nu)\right]^2 
\e
Here $N$ is the total number of antennae
Note that the baseline fraction function $\rho(U,\,\nu)$ is different
for different interferometric arrays and plays an important role in
determining the sensitivity of an array. 
This function is also frequency dependent. 
The signal and the system noise is expected to be isotropic and only depends
on the magnitude of $U$. 
We take average over all cells of area $\Delta U^2$ in the circular
annulus between $U$ and $U+\Delta U$. 
Number of independent elementary bins in a circular annulus between $U$ and
$U+\Delta U$ is $2\pi U \Delta U/ \Delta U^2$. 
Moreover there are $B/\Delta \nu$ independent channels for the total bandwidth 
$B$. 
Combining all these we have 
\be
N_p=\frac{1}{2}\left[ \frac{N(N-1)}{2}\frac{T}{\Delta t}\Delta U^2
  \rho(U,\nu)\right]^2\frac{B}{\Delta \nu} \frac{2\pi U \Delta
  U}{\Delta U^2}
\e
The noise {\it rms} in the visibility correlation can then be written as
\be
\sigma_{VV}=\frac{4}{\sqrt{2\pi} \,
  N(N-1)}\left[\frac{T_{sys}}{K}\right]^2\frac{1}{T\sqrt{\Delta 
\nu B}}\,\frac{1}{U^{0.5}\Delta U^{1.5}\rho(U,\nu)}
\label{eq_ns1}
\e
The bin size $\Delta U$ is determined by the field of view of the antenna and
for a Gaussian antenna beam pattern $A(\theta)=e^{-\theta^2/\theta_0^2}$, one
can show that $\Delta U=1/\pi \theta_0$. 
The equation \ref{eq_ns1} gives the system noise in the visibility correlation
for any radio experiment. 
Note that, the system noise {\it rms} in the visibility correlation scales as
$\sim 1/N(N-1)T$ unlike the noise rms in the image which scales as $\sim
1/\sqrt{N(N-1)T}$. 

There are two effects that we ignore in the analysis here. 
\begin{itemize}
\item
At small $U$, comparable with the inverse of the field of view, cosmic
variance also limits our ability to measure the visibility correlations.  
As the field of view of the GMRT is (much) smaller than the MWA, this concern
is more relevant in that case. 
Cosmic variance is subdominant to the system noise as long as we work with
scales that are much smaller than the field of view, or work with $U$ that are
much larger than the one that corresponds to the field of view.
\item
For MWA, the field of view is very large and one needs to take the curvature
of sky into account \citep{2008MNRAS.389.1163M, 2001PhRvD..63l3001N,
  2005PhRvD..71h3009N}. 
We ignore this as we are not estimating the signal at very large angles.
\end{itemize}

We consider three instruments: the GMRT, MWA and a hypothetical instrument
MWA5000. 
The GMRT has a hybrid distribution with $14$ antennae are distributed randomly
in the central core of area $1$~km~$\times 1$~km and $16$ antennae are placed
along three arms ($Y$ shaped). 
The antennae are parabolic dishes of diameter $45$~m. 
The MWA will have $500$ antennae distributed within a circular region of
radius $750$~m. 
The antennae are expected to follow $\rho_{ant}(r)\propto 1/r^2$ distribution
with a $20$~m flat core. 
The hypothetical instrument MWA5000 is almost similar to the MWA but with
$5000$ antennae. 
We assume antennae are distributed as $\rho_{ant}(r)\propto 1/r^2$ within
$1000$~m radius and with a flat core of radius $80$~m
\citep{2006ApJ...653..815M}. 
The normalized baseline distribution $\rho(U,\nu)$ for frequency $\nu=330$~MHz
is presented in the Figure~\ref{rho} for the three instruments (see
\citet{datta2} for details). 
Table~\ref{table_system_param} lists the wavebands and other instrument
parameters for which analysis is done in this paper.  
We assume the effective antenna collecting area ($A_{eff}$) to be same as the
antenna physical area. 
We also list parameters used for calculation of noise in visibility
correlations.  

\subsection{Noise in Images}
\label{subsec_imagenoise}

Observed visibilities are used to construct an image of the sky for each
frequency channel. 
The size of the image is of the order of the primary beam $\theta_0$ and the
resolution depends on the largest baselines used.
Both the size of the image and resolution, or the pixel size depend on the
frequency of observation in the same manner. 

Unlike visibilities, that are uncorrelated outside of the small bins, pixels
in an image are not completely uncorrelated.  
The correlation of pixels in a raw image is similar to the {\it dirty beam}. 
Therefore the analysis of noise in images as a function of scale is a fairly
complex problem. 
Scaling of noise with bandwidth is easy as the noise for different frequency
channels is uncorrelated.
{\it rms} noise for a given pixel can be computed using (assuming two
polarizations) \citep{2001isra.book.....T} 
\begin{equation}
\sigma_{image} = \frac{T_{sys}}{K} \frac{1}{\sqrt{ \Delta \nu \Delta t}
  \sqrt{2\, N\left(N-1\right)}} 
\label{eqn_imagenoise}.
\end{equation}
The symbols in the above equation have the same meaning as described in
\S{4.2}.  

We find that for the $z=3.3$ window of the GMRT $\sigma_{image} \simeq
3.6$~$\mu$Jy for an integration time of $2000$~hours and a bandwidth of
$1$~MHz. 
This has been computed for the central square of the GMRT with $N=14$.
The corresponding number of $z=1.3$ is $\sigma_{image} \simeq
7.8$~$\mu$Jy for an integration time of $400$~hours and a bandwidth of
$1$~MHz.

As a first approximation, we assume that noise in images is uncorrelated and
that it scales as the square root of the number of pixels over which signal is
smoothed. 
This allows us to estimate signal to noise ratio for extended structures, but
it must be noted that this analysis is approximate.

\section{Prospects for Detection: Simulated Maps And Signal}
\label{sec_detection}

In this section we describe results from simulated \h1 maps as described
in the previous sections.

\subsection{Visibility Correlation}
\label{subsec_viscorr}

We use the power spectrum of the \h1 distribution derived from our N-Body
simulations to compute the visibility correlations. 
For large scales, larger than the simulation size, we patch this with the
linearly extrapolated power spectrum with a constant bias.
This is the only input required in Eqn.~(\ref{eq_vis_cor}) for calculation of
visibility correlation. 

\begin{figure*}
\begin{center}
\begin{tabular}{cc}
\includegraphics[width=1.9truein,angle=270]{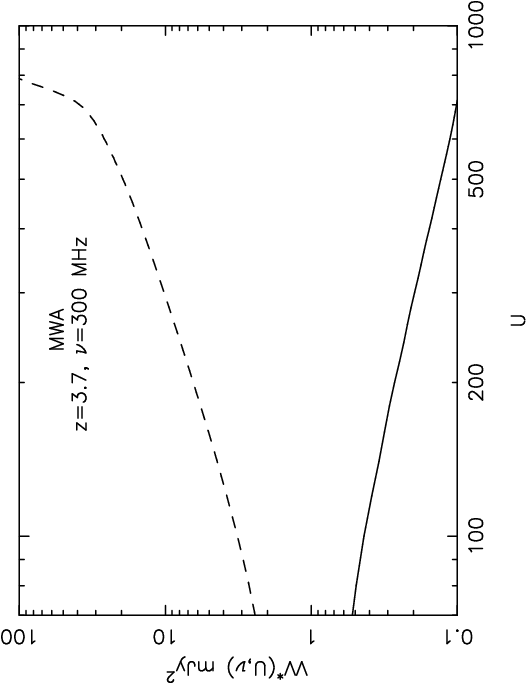} &
\includegraphics[width=1.9truein,angle=270]{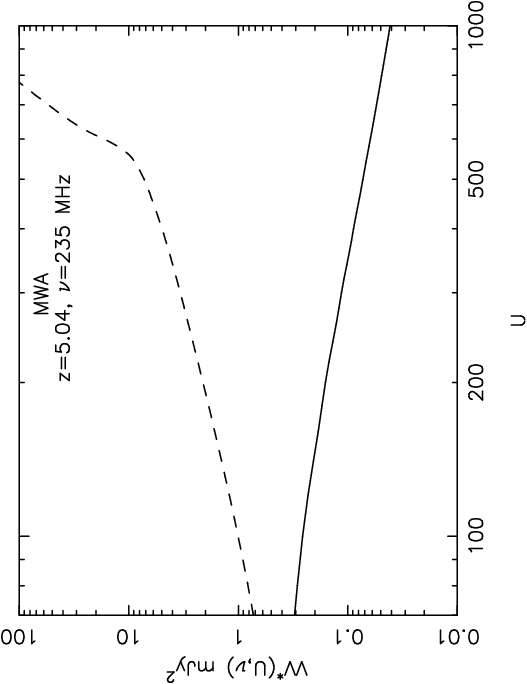} \\
\includegraphics[width=1.9truein,angle=270]{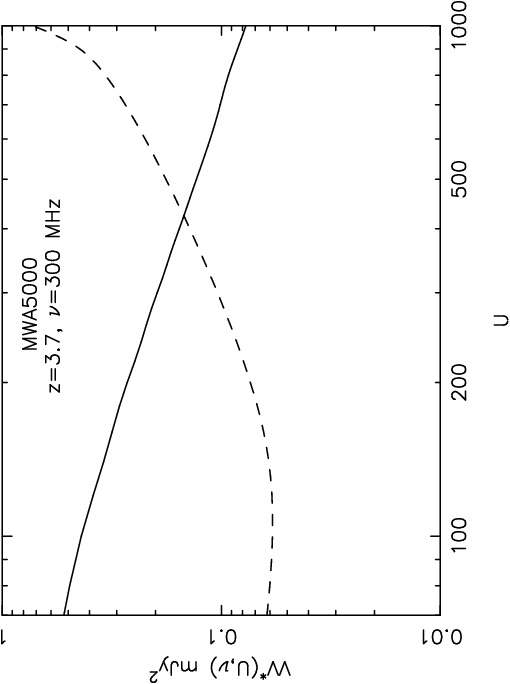} & 
\includegraphics[width=1.9truein,angle=270]{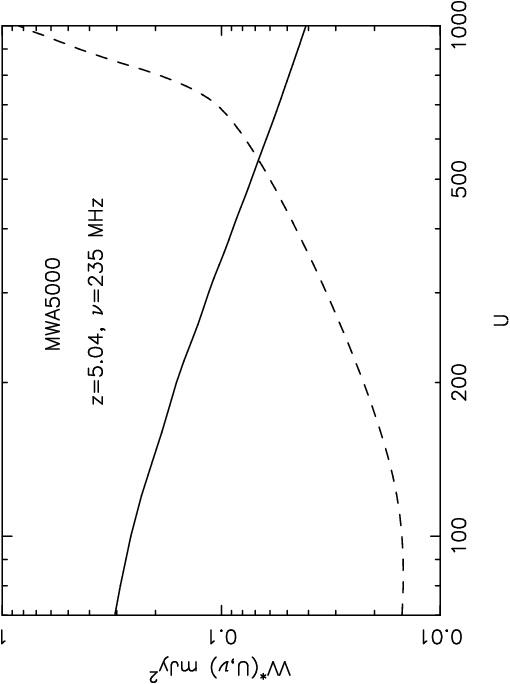} \\
\includegraphics[width=1.9truein,angle=270]{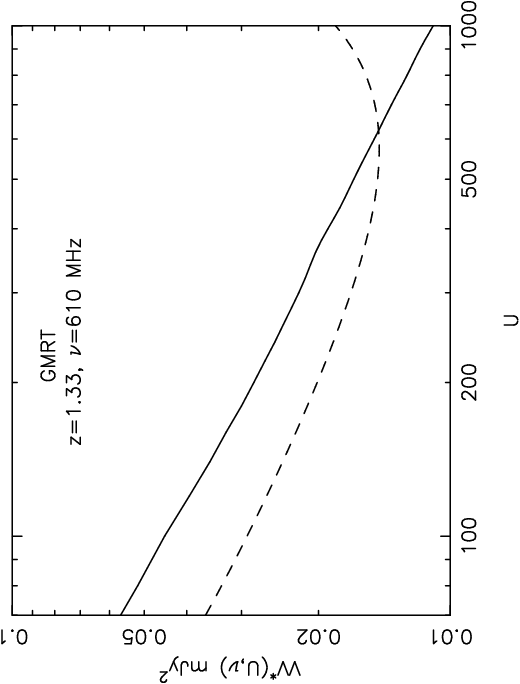} & 
\includegraphics[width=1.9truein,angle=270]{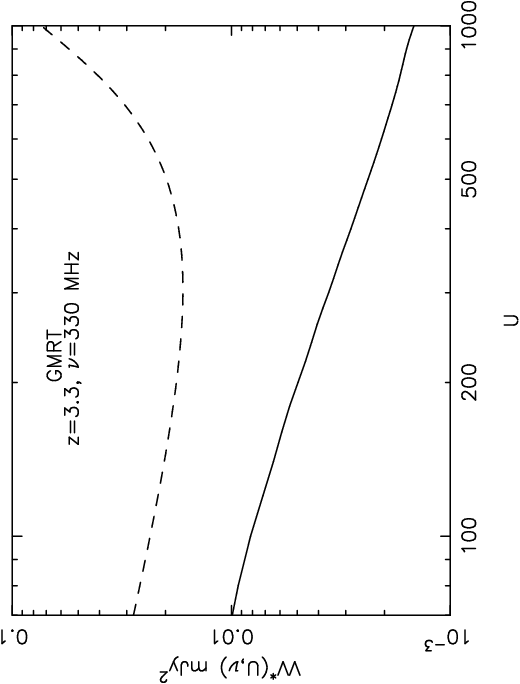} \\ 
\includegraphics[width=1.9truein,angle=270]{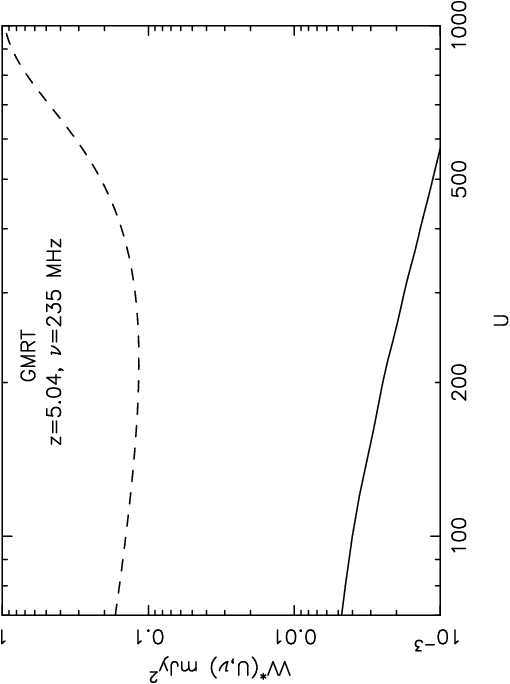} &\\
\end{tabular}
\end{center}
\caption{This figure shows the expected visibility correlation as a function
  of $U$.  The expected signal is shown with a solid curve and the expected
  noise in the visibility correlation is shows by a dashed curve.  This has
  been shown for three instruments: GMRT, MWA
  and a hypothetical instrument MWA5000. The noise
  in each panel has been computed assuming $10^3$ hours of integration time
  and that visibilities are correlated over $1$~MHz. We note that the
  MWA5000 should 
  detect visibility correlations at $U < 500$ at $z \simeq 5$ and at $U < 400$
  at $z \simeq 3.7$.  The detection at smaller $U$ should be possible at a
  high significance level.  Prospects of detection with the GMRT are
  encouraging for $z=1.3$ where it should be possible to make a detection for
  $U < 600$.  It is possible to enhance the signal to noise ratio by combining
  data from nearby bins in $U$, thus detection may be possible in a 
  shorter time scale as well.}
\label{fig_vis_corr}
\end{figure*}

Visibility correlation for various redshifts is shown in 
Figure~\ref{fig_vis_corr}.  
We have also plotted the expected system noise in each of the panels.  
We use the the functional form of $\rho(U, \nu)$ from \citet{datta2}
(also see Figure~\ref{rho}).  
The expected visibility correlation is shown as a function of $U$.  
The expected signal is shown with a solid curve and the expected noise in the
visibility correlation is shows by a dashed curve.  
This has been shown for three instruments: GMRT, MWA
and a hypothetical instrument MWA5000. 
The noise in each panel has been computed assuming $10^3$ hours of integration
time and visibilities correlated over $1$~MHz.  
We find that MWA5000 should detect visibility correlations 
for baselines $U < 500$ at $z \simeq 5$ and 
at $U < 400$ at $z \simeq 3.7$. 
Higher observation time is needed for the MWA to detect the visibility
correlations.  
The detection at smaller $U$ should be possible at high significance 
level. 
Whereas the prospects for detection with the GMRT are 
encouraging for $z=1.3$ where it should be possible to make 
a detection for baselines $U < 600$. 

It is possible to enhance the signal to noise ratio by combining data from
nearby bins in $U$, thus detection may be possible in a shorter time scale as
well.
The GMRT has been in operation for more than a decade, and its
characteristics are well understood.  
Therefore it is important to make an effort to observe \h1 fluctuations at $z
\simeq 1.3$ with the instrument. 
The observations need not pertain to the same field as the region observed by
the GMRT in a single observation is fairly large and we do not expect
significant fluctuations in the \h1 power spectrum and hence the visibility
correlation from one field to another.
Thus good quality archival data for a few fields can be combined, in
principle, to detect visibility correlations. 

Detection of visibility correlations in turn gives us a measurement of the
power spectrum for the \h1 distribution \citep{2001JApA...22..293B}.  
This can be used to constrain various cosmological parameters
\citep{2009PhRvD..79h3538B, 2008arXiv0812.0419V, 2008PhRvL.100p1301L}.  
Recently, it has also been pointed out that the power spectrum can also be
used to estimate the characteristic mass of damped lyman alpha systems (DLAS)
\citep{2008arXiv0804.1624W}.

\begin{figure*}
\begin{center}
\begin{tabular}{cc}
\includegraphics[width=2.6truein,angle=270]{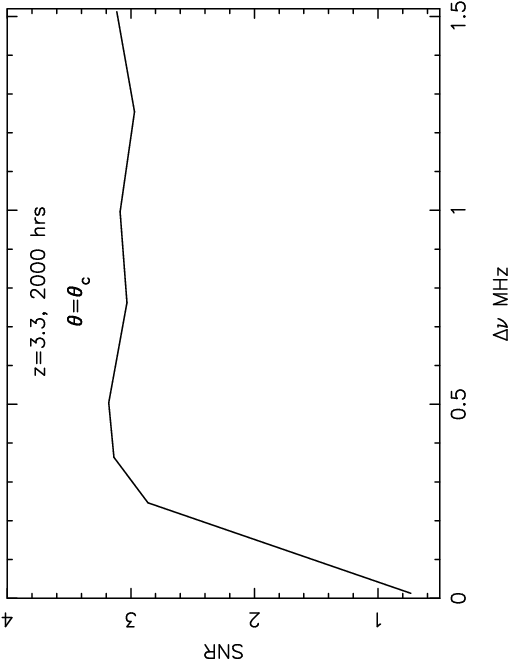} & 
\includegraphics[width=2.6truein,angle=270]{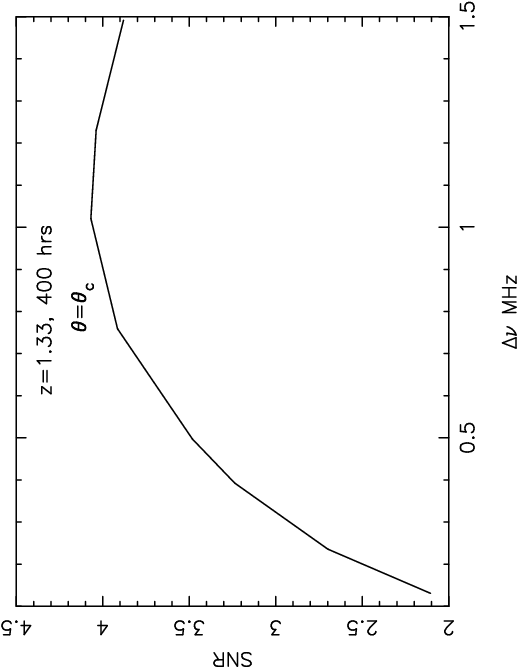} \\
\includegraphics[width=2.6truein,angle=270]{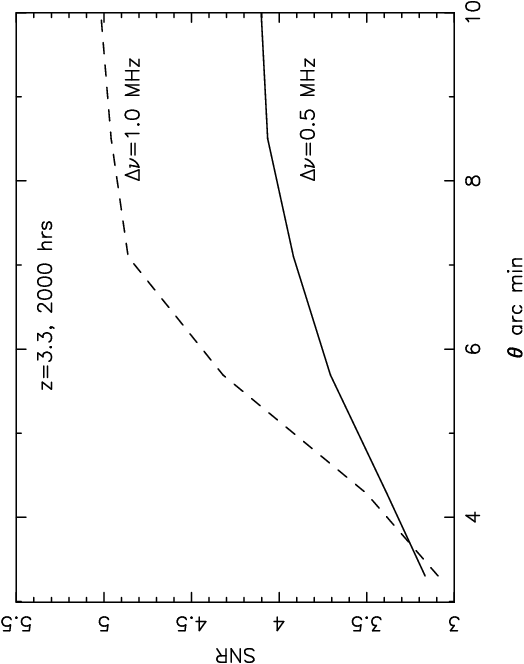} &
\includegraphics[width=2.6truein,angle=270]{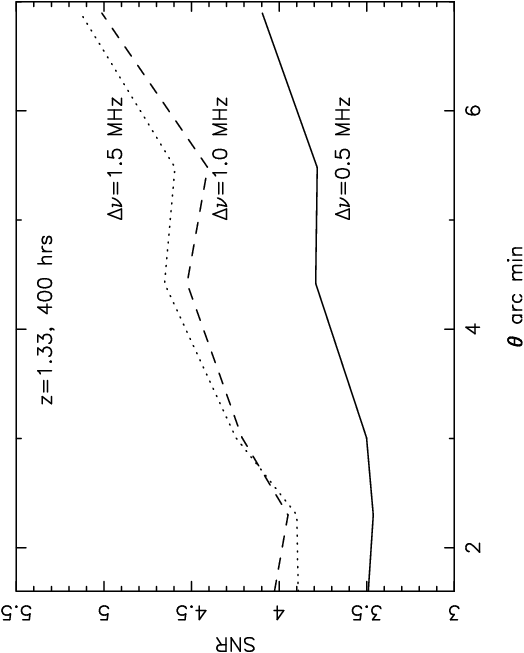} \\
\end{tabular}
\end{center}
\caption{This figure shows the expected signal to noise ratio (SNR) for direct
  detection of rare peaks in the \h1 distribution at high redshifts.  All the
  plots pertain to the GMRT and we use the expected noise in the image for the
  central square.  The left column is for $z \simeq 3.3$ and the right column
  is for $z \simeq 1.3$.  The top row is for variation of the SNR with 
  bandwidth for one pixel, whereas the bottom row shows the variation of SNR
  with the square root of the solid angle over which signal has been smoothed
  for a given bandwidth.  These figures show that a $4-5\sigma$ detection of a
  rare peak is possible in $400$~hours for $z \simeq 1.3$ and in $2000$~hours
  for $z \simeq 3.3$.}
\label{fig_snr}
\end{figure*}

\subsection{Rare Peaks}
\label{subsec_rarepeaks}

We now discuss the possibility of directly detecting rare peaks in the \h1
distribution. 
This is an interesting possibility for instruments like the GMRT and ASKAP
that can resolve small angular scales.
At such scales, non-linear gravitational clustering and the large bias for the
\h1 distribution combine to enhance the amplitude of fluctuations
significantly (\S\ref{sec_results}). 
It is also of interest to see whether this enhancement can make detection of
rare peaks as easy as statistical detection via visibility correlations. 

Figure~\ref{fig_snr} shows the expected signal to noise ratio (SNR) 
for the brightest region in the simulated maps if it is observed using 
the central square of the GMRT. 
The left column is for $z \simeq 3.3$ and the right column is for $z \simeq
1.3$.  
The top row shows variation of the SNR with bandwidth for one pixel, whereas
the bottom panel shows the variation of SNR with the square root of the solid
angle over which signal has been smoothed for a given bandwidth.  
In plots of variation of SNR with the bandwidth, the peak is very close to the
FWHM of spectral lines for large structures (see spectra of rare peaks in
\S\ref{fig_mockmaps}). 
Although SNR falls off towards smaller bandwidths, it does not decline
significantly towards larger bandwidths and this can be attributed to the
presence of nearby lines and is a reflection of clustering at small scales. 

These figures show that a $4-5\sigma$ detection of a rare peak is possible in
$400$~hours for $z \simeq 1.3$ and in $2000$~hours for $z \simeq 3.3$.
The time for detection here is comparable to, indeed slightly less than the
time required for a statistical detection.  
This is a very exciting possibility as it allows us to measure the \h1
fraction of a fairly large region, potentially leading to a measurement of
$\Omega_{\h1}$. 
However, unlike statistical detection of the \h1 distribution, the integration
time cannot be divided across different fields for a direct detection. 

\begin{figure}
\includegraphics[width=3.2truein]{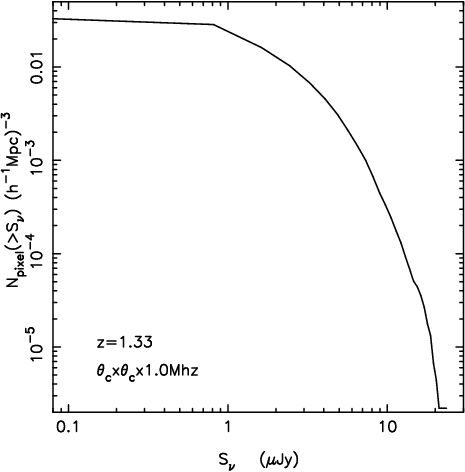}
\caption{This figure shows the number density of pixels in the simulated radio
map with signal above a given threshold.  Here, each pixel corresponds to the
angular resolution of the central square of the GMRT and a bandwidth of
$1$~MHz.}
\label{fig_pixelcount}
\end{figure}

The volume sampled in observation of one field with the GMRT at $z \simeq 3.3$
is much larger than the volume of the simulation used here.  
Thus there is a strong possibility of finding even brighter region in a random
pointing at that redshift. 
On the other hand, the volume sampled by the GMRT at $z \simeq 1.3$ is
smaller by a factor of more than $5$ as compared to the volume of the
simulation used for making mock maps.  
We have studied the distribution function for signal in pixels and find that
there are a sufficiently large number of bright pixels in the simulated map
and a random pointing should have a rare peak comparable to the one discussed
here within the field of view. 
This is shown in Figure~\ref{fig_pixelcount} where we have shown 
the number density of pixels in
the simulated radio map with signal above a given threshold.  Here, each pixel
corresponds to the angular resolution of the central square of the GMRT and a
bandwidth of $1$~MHz.
The volume covered by one GMRT field with a bandwidth of $32$~MHz should
contain more than ten pixels brighter than $10$~$\mu$Jy.  
For comparison, the brightest pixel in the simulated map gives out twice this
flux. 

We do not discuss the possibility of detection of rare peaks with the MWA as
it has a poorer sensitivity at small scales. 

Note that in our results we have ignored
the  foregrounds. The foregrounds at relevant frequencies will dominate
  over the cosmological HI 21-cm signal and needs to be
  subtracted. Line of sight modes $k_{\parallel} \lesssim 0.065 \sqrt{
    \frac{4.5}{1+z}}\left (\frac{8\,{\rm MHz}}{B}\right){\rm Mpc^{-1}}$ can
  not be measured because of foreground subtraction, $B$ is the 
  bandwidth over which foregrounds are subtracted
  \citep{2006ApJ...653..815M}. This would effect the baselines $U\lesssim
  80$ for a ${8\,{\rm MHz}}$ subtraction bandwidth. Rare peaks are
  much smaller objects ($< 10\, {\rm arc min}$) than the scales where
  the foreground subtraction would  affect and hence we 
  expect that our results on rare peaks should not be affected by
  foregrounds.

\section{Discussion}
\label{sec_discussion}

Several attempts have been made in recent years to model the \h1 distribution
in the post-reionization universe \citep{1990MNRAS.247..510S,
  1993MNRAS.265..101S, 1995MNRAS.272..544K, 1997MNRAS.289..671B,
  2001JApA...22...21B, 2001JApA...22..293B, 2003ASPC..289..251B,
  2004JApA...25...67B, 2005MNRAS.356.1519B, 2008PhRvL.100p1301L,
  2008MNRAS.383.1195W, 2008PhRvD..78j3511P, 2009arXiv0912.2130W}, with the
recent resurgence in interest being due to upcoming radio telescopes and
arrays.   
In this paper we have revisited the issue using dark matter simulations and an
ansatz for assigning \h1 to dark matter haloes. 
This is amongst the few attempts made using simulations where the smallest
haloes that may contain significant amount of \h1 are resolved, and, the
simulations are large enough to limit errors due to a finite size of the
simulation box.   

The prospects for detection of the \h1 distribution at high redshifts are very
promising. 
The GMRT can be used to measure visibility correlations at $z \simeq 1.3$.  
We can expect the hypothetical MWA5000 to measure visibility
correlations at $z > 3.7$ and 
provide an estimate of the power spectrum. The integration time
required is not very large. 
This is very encouraging as no other existing or upcoming instrument can probe
this redshift, though ASKAP and MeerKAT will be able to probe the \h1 
distribution at $z < 1$. 
Measurement of visibility correlations, and hence the \h1 power spectrum 
can be used to constrain several cosmological parameters.
If we are able to make measurements at several redshifts then it becomes
possible to constrain models of dark energy. 
Observations of visibility correlations can also be used to put constraints on
the characteristic mass of damped Lyman alpha systems
\citep{2008arXiv0804.1624W}. 
The quadratic dependence of the visibility correlations on $\Omega_{\h1}$ can
be used to determine the density parameter for \h1 at high redshifts. 

An interesting potential application of the method used here is to make
predictions for $z < 1$ where statistical and individual detection 
with the GMRT and ASKAP is likely to be much easier.  
It is obvious that detection of rare peaks of \h1 with the GMRT may be even
easier at lower redshifts ($z \leq 0.4$).  
The main issue to be addressed at these redshifts 
is that the observed volume becomes small
and hence we can expect significant scatter in the observed visibility
correlations from one field to another, and across frequency channels. 
This is true even for rare peaks and 
one may need to observe several fields before finding a very
bright object. Or one may need to observe for longer periods of 
time in order to observe a not so rare peak.
This is particularly relevant for the GMRT where the primary beam covers
a solid angle that is nearly $14$ times smaller than that covered by ASKAP. 
We propose to study this issue in a later publication, where we will also
discuss prospects of observing \h1 using ASKAP and MeerKAT.
We are also working on using wavelet based methods for detection of rare peaks
in the \h1 distribution.  

In this work we have used a fairly simple \h1 assignment scheme while ignoring
the evolution of gas content of haloes. 
We are working on rectifying this shortcoming and expect to use a
semi-analytical model to improve this aspect of modeling.  

Key conclusions of this paper may be summarized as follows:
\begin{itemize}
\item
We find that non-linear gravitational clustering enhances the amplitude of
perturbations by a significant amount at small scales. 
\item
\h1 distribution is strongly biased at high redshifts and this enhances the
\h1 power spectrum significantly as compared to the dark matter power
spectrum.  
Bias $b(k)$ is scale independent at large scales $k \ll k_{nl}$, where
$k_{nl}$ is the scale of non-linearity.
\item
Bias decreases sharply from high redshifts towards low redshifts. 
This leads to a gradual decrease in the brightness temperature power
spectrum. 
The change in the amplitude of the brightness temperature power
spectrum is slow, being less than a factor two between $z=5.1$ and $z=1.3$. 
\item
Brightest regions in the simulated radio maps are found to be extended, with
an angular extent much larger than the resolution of the GMRT. 
This can be used to enhance prospects of detection. 
\item
Spectra in the simulated maps appear to have an FWHM of around $1$~MHz for $z
= 1.3$ and about half of this for $z=3.34$. 
A comparison with earlier work for $z=3.34$ shows that the FWHM is found to be
smaller in the present work. 
We attribute this to the low \h1 fraction assigned to the most massive haloes,
whereas such haloes were not excluded in earlier work.
\item
The MWA5000 should detect visibility correlations at $U < 300$ at $z \simeq 5$
and at $U < 700$ at $z \simeq 3.7$.  The detection at smaller $U$ should 
be possible at a high significance level.
\item
Prospects of detection with the GMRT are encouraging for $z=1.3$ where it
should be possible to make a detection for $U < 600$.
\item
Signal for the brightest source of redshifted $21$~cm radiation from $z=1.3$
appears to be significant, and within reach of an instrument like the GMRT. 
\item
Detection of rare peaks in the \h1 distribution offers exciting
possibilities since it does not require a
much longer integration time as compared to statistical detection.
The size of the region represented in these rare peaks is fairly large, and it
should possible to establish the mass contained in these regions using
observations in other wavebands. 
The direct detection at $z \simeq 1.3$ requires only a few hundred
hours of observing time with an existing instrument.
This may allow us to estimate $\Omega_{\h1}$ in emission, and hence provide
and independent measurement of the amount of cold gas.
\end{itemize}

\section*{Acknowledgments}

Computational work for this study was carried out at the cluster
computing facility in the Harish-Chandra Research Institute
(http://cluster.hri.res.in/).  
This research has made use of NASA's Astrophysics Data System. 
KKD is grateful for financial support from Swedish Research Council
(VR) through the Oscar Klein Centre. The authors would like to thank
the anonymous referee  
for suggestions which helped improve the paper.
The authors would like to thank Jayaram Chengalur, Somnath Bharadwaj, 
Tirthankar Roy Choudhury, Abhik Ghosh and Prasun Dutta 
for useful comments and discussions.

\label{lastpage}

\end{document}